\documentclass[a4paper]{jpconf}
\usepackage{graphicx}
\newcommand{\ba}{\begin{eqnarray}}
\newcommand{\ea}{\end{eqnarray}}

\begin{document}

\title{Electromagnetic and weak decays of baryons in the unquenched quark model}

\author{Roelof Bijker, Gustavo Guerrero-Navarro and Emmanuel Ortiz-Pacheco}
\address{Instituto de Ciencias Nucleares, 
Universidad Nacional Aut\'onoma de M\'exico, 
A.P. 70-543, 04510 M\'exico, D.F., M\'exico}
\ead{bijker@nucleares.unam.mx}

\begin{abstract}
In this contribution, we discuss the electromagnetic and weak decays of baryons in 
the unquenched quark model and show that the observed discrepancies between the 
experimental data and the predictions of the constituent quark model can be accounted 
for in large part by the effects of sea quarks. Finally, the obtained results are 
discussed in terms of flavor-symmetry breaking. 
\end{abstract}

\section{Introduction}

The constituent quark model (CQM) describes the nucleon as a system of three constituent, 
or valence, quarks. Despite the successes of the CQM (e.g. masses, electromagnetic couplings, 
and magnetic moments), there is compelling evidence for the presence of sea quarks from other  
observables such as the observed flavor asymmetry of the proton, the proton spin crisis, 
and the systematics of strong decays of baryons.  

In the CQM, baryons are described in terms of a configuration of 
three constituent (or valence) quarks neglecting the effects of pair-creation (or 
continuum couplings). Above threshold these couplings lead to strong decays and below 
threshold to virtual higher-Fock components (such as $qqq-q\bar{q}$) in the baryon wave function. 
The effects of these multiquark configurations (or sea quarks) is studied by unquenching 
the CQM. 

In this contribution, we study the importance of sea quarks for the electromagnetic 
and weak couplings of baryons. 

\section{Unquenched quark model}

In the unquenched quark model (UQM), the effects of sea quarks are included via a $^{3}P_{0}$ 
quark-antiquark pair-creation mechanism \cite{tornqvist,zenc,baryons,uqm1,uqm2}. The pair-creation 
mechanism is inserted at the quark level and the one-loop diagrams are calculated by summing 
over a complete set of intermediate baryon-meson states. As a result, the baryon wave function is 
given by a superposition of a valence contribution and higher-Fock components consisting of 
intermediate baryon-meson configurations  
\ba 
\mid \psi_A \rangle = {\cal N}_A \left\{ \mid A \rangle + \gamma 
\sum_{BC l J} \int d \vec{K} k^2 dk \, \mid BC,l,J; \vec{K},k \rangle \, 
\frac{ \langle BC,l,J; \vec{K},k \mid T^{\dagger} \mid A \rangle } 
{M_A - E_B(k) - E_C(k)} \right\} ~.
\label{wf}
\ea
Here $\vec{k}$ and $l$ denote the relative radial momentum and relative orbital angular momentum of 
the baryon-meson system BC. The operator $T^{\dagger}$ creates a quark-antiquark pair in the 
$^{3}P_0$ state with the quantum numbers of the vacuum \cite{uqm1,uqm2,roberts} 
\ba
T^{\dagger}( ^3P_0 ) &=& -3 \gamma \,\int d \vec{p}_4 \, d \vec{p}_5 \, 
\delta(\vec{p}_4 + \vec{p}_5) \, C_{45} \, F_{45} \,  
{e}^{-\alpha_d^2 (\vec{p}_4 - \vec{p}_5)^2/8 }\, 
\nonumber\\
&& \hspace{1cm} \left[ \chi_{45} \, \times \, {\cal Y}_{1}(\vec{p}_4 - \vec{p}_5) \right]^{(0)}_0 \, 
b_4^{\dagger}(\vec{p}_4) \, d_5^{\dagger}(\vec{p}_5) ~.
\label{3P0} 
\ea
The quark-antiquark pair is characterized by a color singlet wave function $C_{45}$, 
a spin triplet wave function $\chi_{45}$ with spin $S=1$ and a solid spherical harmonic 
${\cal Y}_{1}(\vec{p}_4 - \vec{p}_5)$ that indicates that the  quark and antiquark are 
in a relative $P$ wave. $F_{45}$ denotes the flavor wave function of the created quark-antiquark pair 
\cite{F45,Jacopo}
\ba
\frac{1}{\sqrt{2+(m_n/m_s)^2}} \left[ | u\bar{u}\rangle + |d\bar{d}\rangle + \frac{m_n}{m_s} |s\bar{s}\rangle \right] ~,
\label{qqbar}
\ea
which, in the limit of equal quark masses, reduces to the usual expression for a flavor singlet. 
The ratio of nonstrange to strange quark masses is determined from the quark magnetic moments as 
\ba
\frac{m_n}{m_s} = \frac{m_u+m_d}{2m_s} = \frac{\mu_s(\mu_u-2\mu_d)}{2\mu_u\mu_d} ~.
\ea

For some observables like the magnetic moments, the effects of quark-antiquark pairs to a large 
extent can be taken into account by introducing effective (or renormalized) values of the model 
parameters \cite{uqm1}. As a result, the CQM results are not altered much by the introduction of 
higher Fock components in baryon (and meson) wave functions. There are other observables for which 
the explicit inclusion of sea quarks is crucial since the quark-antiquark pairs provide non-neglible contributions that cannot be accounted for by renormalization of model parameters. Examples of the 
latter are the orbital angular momentum in the spin of the 
proton \cite{uqm1}, the flavor asymmetry of the proton \cite{uqm2}, strangeness content of 
nucleon electromagnetic form factors \cite{baryons,uqm3}, strangeness suppression \cite{uqm4}, 
and self-energy corrections to baryon and meson masses \cite{uqm5,FS}. In the next section, we discuss 
the importance of higher-Fock components in electromagnetic and weak decays of baryons. 

\section{Results}

In this section, we discuss some recent results for electromagnetic and weak couplings. 
A more detailed account will be given in future publications \cite{gustavo,emmanuel}. 
In the present calculation, the sum over intermediate states is limited to octet and decuplet 
baryons in combination with pseudoscalar octet and singlet mesons. 
The contributions of radially excited baryons and mesons are not taken into account.  

\subsection{Electromagnetic decays}

The experimental data obtained by the CLAS Collaboration for the electromagnetic decays 
of $\Sigma$ hyperons of the baryon decuplet show a large deviation from the CQM predictions 
\cite{Keller1,Keller2}. Table~\ref{emtab} shows that the experimental widths are underpredicted 
by almost a factor of two. Here we study the effect of sea quarks for the electromagnetic 
decays of the decuplet baryons.  

The radiative width for this process can be expressed in terms of the 
transition magnetic moment $\mu_{AB}$
\ba
\Gamma(A \rightarrow B \gamma) &=& \frac{\alpha E_B p_{\gamma}^3}{2m_A m_N^2} \mu_{AB}^2 ~.
\ea
The results in Table~\ref{emtab} show that for the $\Delta$ resonance the coupling to the 
pseudoscalar meson cloud (mostly pions \cite{Salamanca}) accounts in large part for the observed 
discrepancy between the quark model value $399$ keV and the experimental value $704 \pm 63$ keV. 
However, for the $\Sigma$ hyperons the calculated width is larger than the CQM value 
but still rather far from the experimental result.  

\begin{table}[h]
\centering
\caption{Electromagnetic decay widths in keV.} 
\label{emtab}
\vspace{5pt}
\begin{tabular}{lccc}
\hline
\noalign{\smallskip}
& CQM & UQM & Exp \cite{PDG} \\
\noalign{\smallskip}
\hline
\noalign{\smallskip}
$\Gamma(\Delta \rightarrow N \gamma)$                 & 399 & 606 & $704 \pm 63$ \\
$\Gamma(\Sigma^{\ast 0} \rightarrow \Lambda \gamma)$  & 260 & 318 & $451 \pm 77$ \\
$\Gamma(\Sigma^{\ast +} \rightarrow \Sigma^+ \gamma)$ & 110 & 131 & $254 \pm 59$ \\
\noalign{\smallskip}
\hline
\end{tabular}
\end{table}

\subsection{Weak decays}

Semi-leptonic decay processes of baryons are described by means of the axial couplings. 
In the case of octet baryons they can be expressed in terms of the couplings $F$ and $D$. 
In the quark model their values are given 
by $F=2/3$ and $D=1$. In the Cabibbo approach, $F$ and $D$ are determined from the experimental 
axial couplings, $g_A(n \rightarrow p)$ and $g_A(\Sigma^- \rightarrow n)$, leading to effective 
values, $F=0.465$ and $D=0.805$. With these values the semileptonic decay processes of baryons 
are described very well \cite{Cabibbo}. 

In Table~\ref{weaktab} we show a comparison of the results for the axial couplings in the CQM 
and the Cabibbo approach with those of the unquenched quark model. The contribution of the 
pseudoscalar mesons (and especially that of the pions \cite{Salamanca}) in the UQM is responsable 
for a substantial lowering of the neutron axial coupling from the CQM value thus bringing it in 
much closer agreement with experiment without the need to introduce effective values of $F$ and $D$. 
On the other hand, the result for the $\Sigma^-$ hyperon which is described very well in the CQM 
is hardly changed. Table~\ref{weaktab} shows that the effective values of $F$ and $D$ used in the 
Cabibbo approach to a large extent can be accounted for in the UQM by the coupling to the 
pseudoscalar mesons. A similar conclusion was reached in an earlier study of the meson-cloud model \cite{Speth}. 

\begin{table}[h]
\centering
\caption{Axial couplings for $\beta$ decays.}
\label{weaktab}
\vspace{5pt}
\begin{tabular}{lcrrrcc}
\noalign{\smallskip}
\hline
\noalign{\smallskip}
& $g_A$ & CQM & UQM & Cabibbo && Exp \cite{PDG} \\
\noalign{\smallskip}
\hline
\noalign{\smallskip}
$n \rightarrow p$ & $F+D$ & $1.67$ & $1.35$ & $1.27$ & * & $1.2701 \pm 0.0025$ \\
\noalign{\smallskip}
$\Sigma^- \rightarrow n$ & $F-D$ & $-0.33$ & $-0.31$ & $-0.34$ & * & $-0.340 \pm 0.017$ \\
\noalign{\smallskip}
$\Xi^0 \rightarrow \Sigma^+$ & $F+D$ & $1.67$ & $1.33$ & $1.27$ && $1.21 \pm 0.05$ \\
\noalign{\smallskip}
$\Lambda^0 \rightarrow p$ & $\frac{1}{\sqrt{6}}(3F+D)$ & $1.22$ & $0.93$ & $0.90$ && $0.879 \pm 0.018$ \\
\noalign{\smallskip}
$\Sigma^- \rightarrow \Lambda^0$ & $\sqrt{\frac{2}{3}} D$ & $0.82$ & $0.71$ & $0.66$ && $0.60 \pm 0.03$ \\
\noalign{\smallskip}
$\Xi^- \rightarrow \Lambda^0$ & $\frac{1}{\sqrt{6}}(3F-D)$ & $0.41$ & $0.30$ & $0.24$ && $0.31 \pm 0.06$ \\
\noalign{\smallskip}
\hline
\noalign{\smallskip}
& $F$ & $\frac{2}{3} \;$ & $\frac{2}{3} \;$ & $0.465$ && \\
& $D$ & $1 \;$           & $1 \;$           & $0.805$ && \\
\noalign{\smallskip}
\hline
\end{tabular}
\end{table}

\section{Flavor-symmetry breaking}

Finally, we discuss the UQM results for electromagnetic and weak decays in terms 
of flavor-symmetry breaking effects. In the UQM, the flavor symmetry is broken by the use of the 
physical masses of baryons and mesons in the energy denominator in Eq.~(\ref{wf})  
and the ratio of nonstrange to strange quark masses in the created quark-antiquark pair of 
Eq.~(\ref{qqbar}). The third column of Table~\ref{fsb} shows the results for the $SU(3)$ 
flavor symmetry limit. In this case, the transition magnetic moments depend on a single 
coupling which is normalized to $\mu_{\Delta N}$. For the $\beta$ decays there are two 
independent couplings. Table~\ref{fsb} shows that, whereas for the weak decays the flavor 
symmetry is broken by less than 5 \%, for the transition magnetic moments the breaking is 
much larger $\sim 15 - 25$ \%.  
 
\begin{table}
\centering
\caption{Flavor-symmetry breaking in transition magnetic moments (top) 
and axial couplings (bottom).}
\label{fsb}
\vspace{5pt}
\begin{tabular}{lrrccl}
\hline
\noalign{\smallskip}
& UQM & SU(3) && Exp & \\
\noalign{\smallskip}
\hline
\noalign{\smallskip}
$\mu_{\Delta N}$ & $3.29$ & $3.29$ & $*$ & $3.53 \pm 0.16$ & $\mu_N$ \\
\noalign{\smallskip}
$\mu_{\Sigma^{*+} \Sigma^+}$ & $-2.52$ & $-3.07$ && $-3.50 \pm 0.43$ & $\mu_N$ \\
\noalign{\smallskip}
$\mu_{\Sigma^{*0} \Lambda^0}$ & $2.52$ & $2.85$ && $3.03 \pm 0.27$ & $\mu_N$ \\
\noalign{\smallskip}
\hline
\noalign{\smallskip}
$g_A(n \rightarrow p)$ & $1.35$ & $1.35$ & $*$ & $1.2701 \pm 0.0025$ & \\
\noalign{\smallskip}
$g_A(\Sigma^- \rightarrow n)$ & $-0.31$ & $-0.31$ & $*$ & $-0.340 \pm 0.017$ & \\
\noalign{\smallskip}
$g_A(\Xi^0 \rightarrow \Sigma^+)$ & $1.33$ & $1.35$ && $1.21 \pm 0.05$ & \\
\noalign{\smallskip}
$g_A(\Lambda^0 \rightarrow p)$ & $0.93$ & $0.97$ && $0.879 \pm 0.018$ & \\
\noalign{\smallskip}
$g_A(\Sigma^- \rightarrow \Lambda^0)$ & $0.71$ & $0.68$ && $0.60 \pm 0.03$ & \\
\noalign{\smallskip}
$g_A(\Xi^- \rightarrow \Lambda^0)$ & $0.30$ & $0.29$ && $0.31 \pm 0.06$ & \\
\noalign{\smallskip}
\hline
\end{tabular}
\end{table}

\section{Summary and conclusions}

In this contribution, we studied the importance of higher-Fock components (or sea quarks) 
in electromagnetic and weak decays of baryons in the framework of the unquenched quark model. 
It was shown that the observed discrepancies between the experimental data and the predictions 
of the CQM can be accounted for in large part by the contributions of quark-antiquark pairs 
in the UQM. Moreover, it was found that the effects of flavor-symmetry breaking in the UQM 
are much larger for electromagnetic decays than they are for weak decays.

\ack
This work was supported in part by grant IN109017 from DGAPA-UNAM, Mexico
and grant 251817 from CONACyT, Mexico.


\section*{References}

\begin{thebibliography}{99}

\bibitem{tornqvist}
T\"ornqvist N A 1985 \emph{Acta Phys. Polon. B} {\bf 16} 503

\bibitem{zenc}
Zenczykowski P 1986 \emph{Ann. Phys. (N.Y.)} {\bf 169} 453

\bibitem{baryons}
Geiger P and Isgur N 1997 \emph{Phys. Rev. D} {\bf 55} 299 

\bibitem{uqm1}
Bijker R and Santopinto E 2009 \emph{Phys. Rev. C} {\bf 80} 065210 

\bibitem{uqm2} 
Santopinto E and Bijker R 2010 \emph{Phys. Rev. C} {\bf 82} 062202(R)  

\bibitem{roberts}
Roberts W and Silvestre-Brac B 1992 \emph{Few-Body Systems} {\bf 11} 171		
	 
\bibitem{F45}
Kalashnikova Y S 2005 \emph{Phys. Rev. D} {\bf 72} 034010

\bibitem{Jacopo}
Ferretti F, Galat\`a G, Santopinto E and Vassallo A 2012 
\emph{Phys. Rev. C} {\bf 86} 015204

\bibitem{uqm3}
Bijker R, Ferretti J and Santopinto E 2012  
\emph{Phys. Rev. C} {\bf 85} 035204.

\bibitem{uqm4} 
Santopinto E, Bijker R and Garc{\'{\i}}a-Tecocoatzi H 2016  
\emph{Phys. Lett. B} {\bf 759} 214 
 
\bibitem{uqm5}
Garc{\'{\i}}a-Tecocoatzi H, Bijker R, Ferretti J and Santopinto E 2017  
\emph{Eur. Phys. J. A} {\bf 53} 115

\bibitem{FS}
Ferretti J and Santopinto E 2014 
\emph{Phys. Rev. D} {\bf 90} 094022

\bibitem{gustavo}
Guerrero-Navarro G and Bijker R, to be published. 

\bibitem{emmanuel}
Ortiz-Pacheco E and Bijker R, to be published. 

\bibitem{Keller1}
Keller D {\it et al.} 2011 {\it Phys. Rev. D} {\bf 83} 072004

\bibitem{Keller2}
Keller D {\it et al.} 2012 {\it Phys. Rev. D} {\bf 85} 052004

\bibitem{PDG}
Patrignani C {\it et al.} (Particle Data Group) 2016 
\emph{Chin. Phys. C} {\bf 40} 100001

\bibitem{Salamanca}
Bijker R, Guerrero-Navarro G and Ortiz-Pacheco E 2017  
PoS (Hadron2017) 053 

\bibitem{Cabibbo}
Cabibbo N, Swallow E C and Winston R 2003  
\emph{Annu. Rev. Nucl. Part. Sci.} {\bf 53} 39

\bibitem{Speth}
Holtmann H, Szczurek A and Speth J 1996
\emph{Nucl. Phys. A} {\bf 596} 631

\end{thebibliography}
\end{document}